\makeatletter
  
  \newcommand{\Rmnum}[1]{\expandafter\@slowromancap\romannumeral #1@}
\makeatother
\documentclass{PoS}

\title{The study of $D_{s1}(2460)$ and $D_{s1}(2536)$ mixing}

\ShortTitle{The study of $D_{s1}(2460)$ and $D_{s1}(2536)$ mixing}

\author{\speaker{Xiao-Gang Wu}\\
Institute of High Energy Physics, Chinese Academy of Sciences, Beijing 100049, China\\
        E-mail: \email{wuxiaogang@ihep.ac.cn}}

\author{Qiang Zhao\\
Institute of High Energy Physics, Chinese Academy of Sciences, Beijing 100049, China\\
and  Theoretical Physics Center for Science Facilities, CAS, Beijing
100049, China\\
        E-mail: \email{zhaoq@ihep.ac.cn}}

\abstract{ In this proceeding, we report our recent study of the
mixing mechanism between $D_{s1}(2460)$ and $D_{s1}(2536)$. On the
basis of Godfrey-Isgur model, we consider the $D^*K$ hadron loop
effect on the ${}^3P_1$ and ${}^1P_1$ $c\bar{s}$ states. We
construct the propagator matrix of these two-state system, from
which we can extract the poles as well as the mixing angles. Through
this method, we simultaneously determine the masses, widths and
mixing angle of these two physical states and the results agree well
with the experimental measurement. Besides the mass shift, we also
find that the hadron loop effects can cause a significant shift for
the mixing angles from the value determined in the heavy quark
symmetry limit. }

\FullConference{Sixth International Conference on Quarks and Nuclear Physics\\
         April 16-20, 2012\\
         Ecole Polytechnique, Palaiseau,  Paris}

\begin{document}

\section{Introduction}

Last year, BaBar Collaboration updated their measurement of the mass
and width of $D_{s1}(2536)$ meson~\cite{Lees:2011um}. Before this
measurement, only an up limit of the width had been
set~\cite{Nakamura:2010zzi}. This new measurement should be useful
for the study of the internal structures of $D_{s1}(2536)$ and its
axial-vector partner $D_{s1}(2460)$. Thus, it initiates a lot of
theoretical interests in the mixing of these two states.

We recall first some observations concerning the $P$-wave $D_s$
state. Theoretically, the Godfrey-Isgur (GI)
model~\cite{Godfrey:1985xj} has made a great success in providing an
overall description of charmonium and charmed meson spectra.
However, it is also realized that there exist apparent discrepancies
between theory and experiment when a state is located close to an
open threshold. In the $D_s$ spectrum, the lowest $S$-wave states
$D_s$ and $D_s^*$ can be well explained by the GI model. But for
those four $P$-wave states, i.e. $D_{s0}(2317)$, $D_{s1}(2460)$,
$D_{s1}(2536)$, and $D_{s2}(2573)$, there are significant
discrepancies between the quark model prediction and experimental
observation. For instance, the masses of $D_{s0}(2317)$ and
$D_{s1}(2460)$ are about 100 MeV lower than the quark model
prediction, and a dynamical reason for such large mass shifts may be
due to the $DK$ and $D^*K$ thresholds,
respectively~\cite{Simonov:2004ar,Badalian:2007yr,Coito:2011qn,Zhou:2011sp}.

State mixing makes these two axial-vector states $D_{s1}(2460)$ and
$D_{s1}(2536)$ more interesting. Such a heavy-light quark-antiquark
system could be an ideal place to test the heavy quark symmetry.
Namely, in the heavy quark symmetry limit, $D_{s1}(2460)$ is a pure
$j=1/2$ state and couples to $D^* K$ channel through an $S$ wave,
while $D_{s1}(2536)$ is a pure $j=3/2$ state and couples to $D^* K$
channel through a $D$ wave. The heavy quark symmetry is expected to be
broken at the order of $1/m_c$. Explicit introduction of a broken
mechanism for the heavy quark symmetry in these two axial-vector
states is thus necessary.

Since both $D_{s1}(2460)$ and $D_{s1}(2536)$  are not charge
conjugation eigenstates, and they both have strong $S$-wave
couplings to $D^*K$, $D_s^*\eta$, and $D K^*$, we can then assume
$D_{s1}(2460)$ and $D_{s1}(2536)$ to be mixed states of the pure
${}^3P_1$ and ${}^1P_1$ $c\bar{s}$ states through intermediate meson
loops. We mention that the $D^*K$ channel plays a dominant role in
such a mechanism. By calculating the intermediate meson loop
transitions, we can simultaneously determine the masses, widths and
mixing angles of the two physical states. A similar mechanism has
been applied to the $a_0(980)-f_0(980)$ mixing~\cite{Wu:2007jh}.

We mention that many other scenarios are also proposed to explain
the experimental data for $D_{s1}(2460)$ and $D_{s1}(2536)$, such as
$D^*K$ molecule, tetra-quark state and dynamically generated
state~\cite{Close:2005se,Barnes:2003dj,Faessler:2007us,Guo:2006rp}.
Besides the mass shift, the hadron loop effect as a general
mechanism can also affect the decay and production process. For
instance, it has been used to evade the helicity selection rule in
$\chi_{c1} \to VV$ and $\chi_{c2} \to VP$~\cite{Liu:2009vv}, account
for the large branch ratio of the non-$D\bar{D}$ decay of
$\psi(3770)$~\cite{Zhang:2009kr,Liu:2009dr}, explain the large
isospin violation process $\eta(1405/1475) \to 3\pi$
~\cite{Wu:2011yx}, and predict a direct measurement of open
threshold effects in $e^+ e^- \to J/\psi \pi^0$~\cite{Wang:2011yh}.

\section{Mixing through intermediate meson loops}

In this part, we show our basic formulas for the mixing scheme
through intermediate meson loops. If two states $|a\rangle$ and
$|b\rangle$ can transit to each other through hadron loops, the
propagator of such a two-state system can be written as a $2 \times
2$ matrix $G_{ab}$~\cite{Wu:2011yb}. In general, the propagator
matrix has two poles in the complex energy plane, which correspond
to the physical states. From the pole position, we can determine the
masses and widths of the physical states. The physical states
$|A\rangle$ and $|B\rangle$ should be a mixture of $|a\rangle$ and
$|b\rangle$,
\begin{equation}\label{eqn1}
    \left(\begin{array}{c}
            |A\rangle \\
            |B\rangle
          \end{array}
    \right)
    =
    \left(
    \begin{array}{cc}
      \cos \theta              & - \sin \theta e^{i \phi}\\
      \sin \theta e^{-i \phi}   & \cos \theta
    \end{array}
    \right)
    \left(\begin{array}{c}
            |a\rangle \\
            |b\rangle
          \end{array}
    \right)
    =
    R(\theta,\phi)
    \left(\begin{array}{c}
            |a\rangle \\
            |b\rangle
          \end{array}
    \right)
\end{equation}
where $\theta$ is the mixing angle, and $\phi$ is a possible
relative phase between $|a \rangle$ and $|b\rangle$. Then the
propagator matrix of $|A\rangle$ and $|B\rangle$ becomes $G_{AB}=R
G_{ab} R^{\dagger}$ and should be diagonal. So we can obtain the
mixing parameters $(\theta,\phi)$ by diagonalizing $G_{ab}$.

Here, we show the propagator matrix.  For the spin-0 states, i.e.
scalar or pseudo-scalar, the propagator matrix is
\begin{equation}\label{eqn2}
    G_{ab}=\frac{1}{D_a D_b-D_{ab}^2} \left(\begin{array}{cc}
                                            D_b & D_{ab} \\
                                            D_{ba} & D_a
                                          \end{array}
    \right) \ ,
\end{equation}
where $D_a$ and $D_b$ are the denominators of the single propagators
of $|a\rangle$ and $|b\rangle$, respectively, and $D_{ab}$ is the
mixing term. For the spin-1 states, i.e. vector and axial-vector,
the propagator matrix is
\begin{equation}\label{eqn3}
    G_{ab}^{\mu \nu}=iP^{\mu \nu}
    \frac{\bar{G}_{ab}(s)}{\det{\bar{G}_{ab}}(s)}+i Q^{\mu \nu} \frac{G^L_{ab}}{\det{G^L_{ab}}} \ ,
\end{equation}
where $P^{\mu \nu}\equiv g^{\mu \nu}-p^{\mu}p^{\nu}/p^2$ and $Q^{\mu
\nu}\equiv p^{\mu}p^{\nu}/p^2$ are the transverse and longitudinal
projector, respectively. The poles and mixing angles are only
related to the transverse part and
\begin{equation}\label{eqn4}
    \bar{G}_{ab}(s)\equiv M_{ab}^2-\delta_{ab} s=
    \left(\begin{array}{cc}
       m_b^2+\Pi_b(s)-s     & -\Pi_{ab}(s) \\
       -\Pi_{ab}(s)         & m_a^2+\Pi_a(s)-s
    \end{array}\right) \ ,
\end{equation}
where $\Pi_a,\,\Pi_b$ and  $\Pi_{ab}$ are the transverse
coefficients of the self-energy functions and the mixing term,
respectively.

For the case of $D_{s1}(2460)$ and $D_{s1}(2536)$, we apply the
following mixing scheme
\begin{eqnarray}\label{eqn5}
  |D_{s1}(2460)\rangle &=& \cos\theta | {}^3P_1 \rangle - \sin\theta e^{i \phi} | {}^1P_1 \rangle  \nonumber \\
  |D_{s1}(2536)\rangle &=& \sin\theta e^{-i\phi} | {}^3P_1 \rangle + \cos\theta | {}^1P_1
  \rangle \ .
\end{eqnarray}
In the heavy quark limit, the ideal mixing angle is
$\theta_0=35.26^\circ$ in our convention. Considering parity
conservation and the OZI rule, the important intermediate states
that can couple to $D_{s1}(2460)$ and $D_{s1}(2536)$ are $D^* K$,
$D_{s}^{*} \eta$ and $D K^*$. To compute these diagrams, we need to
know the couplings of the vertices as well as form factors to remove
the ultraviolet (UV) divergences in the loop integrals.

\section{Result and discussion}

At hadronic level, the Axial-vector-Vector-Pseudoscalar vertex can
be written as $i(g_S g^{\mu \nu} + g_D p^\mu p^\nu)$, where $g_S$
and $g_D$ are the $S$ and $D$-wave couplings. Near threshold, the
$D$-wave couplings are suppressed due to the small momentum. So we
only consider the $S$-wave couplings, and use the chiral quark model
to evaluate
$g_S$~\cite{Manohar:1983md,Riska:2000gd,Zhong:2008kd,Zhong:2009sk}.
In the chiral quark model, the light pseudo-scalars and light
vectors are treated as chiral fields.

\begin{figure}[htbp]
\begin{minipage}[l]{0.45\textwidth}
  \centering
  \includegraphics[scale=0.77]{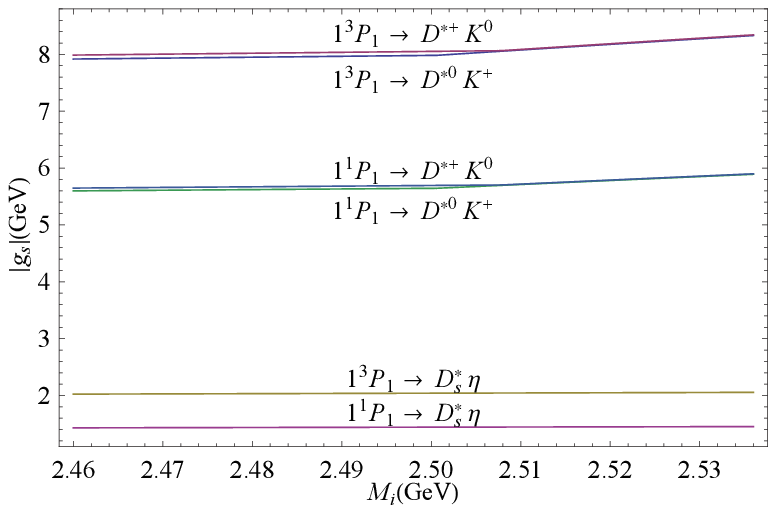}\\
  \vspace{0cm}
  \caption{The absolute values of coupling $g_S$ as a function of the initial meson mass $M_i$~\cite{Wu:2011yb}.}
  \label{fig5}
\end{minipage} \ \ \
\begin{minipage}[l]{0.45\textwidth}
  \centering
  \vspace{-0.1cm}
  \includegraphics[scale=0.7]{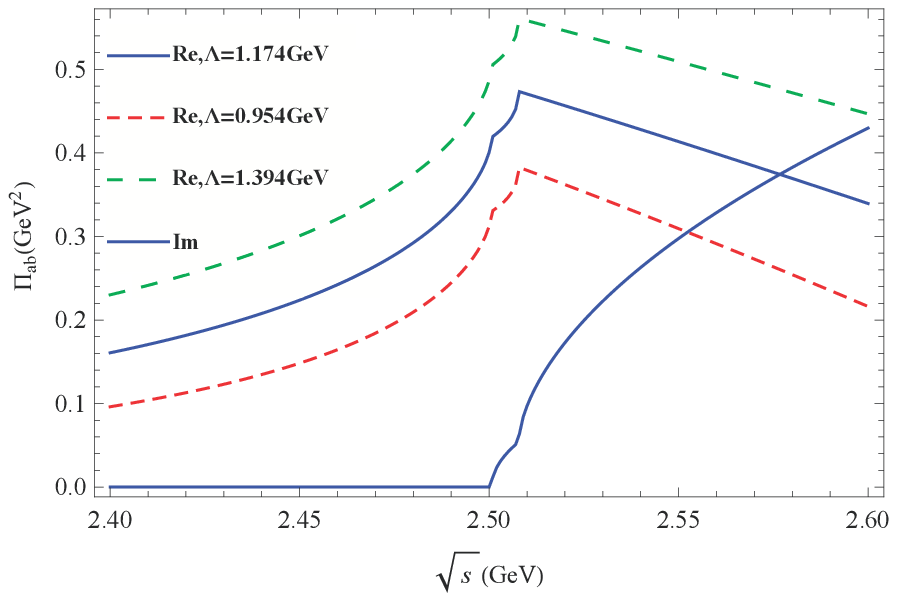}\\
  \vspace{-0.1cm}
  \caption{The mixing term $\Pi_{ab}$ with different cutoff values for the exponential form factor.}
  \label{fig7}
\end{minipage}
\end{figure}

Our result for the $S$-wave couplings are shown in Fig.~\ref{fig5}.
We can see that they are insensitive to the initial meson mass. The
couplings to $D^*K$ are strong, the couplings to $D^*\eta$ are
rather weak, and the couplings to $DK^*$ are vanishing as a leading
approximation. To remove the UV divergences in the loop integrals,
we modify the non-relativistic (NR) exponential form factors in the
quark model to a covariant form, i.e. $\exp(-{q_1^2}/{4 \alpha^2})
\rightarrow \exp({q^2-m^2}/{\Lambda^2})$, where $\Lambda$ is the
cutoff energy and can be fixed by the chiral quark model.

In the propagator matrix Eqs.~(\ref{eqn3}-\ref{eqn4}), what we need
to know are the self-energy functions $\Pi_a$ and  $\Pi_b$, and the
mixing term $\Pi_{ab}$.  They are found to be proportional to each
other~\cite{Wu:2011yb}. In Fig.~\ref{fig7} we show the energy
dependence of $\Pi_{ab}$. The blue lines are the results with cutoff
$\Lambda=1.174 \ \textrm{GeV}$ fixed by the chiral quark model. We
can see that both the real and the imaginary part have two kinks due
to the charged and neutral $D^*K$ thresholds. Below the thresholds
the imaginary part is zero, while above the thresholds it increases
quickly. At the energy near $2460 \ \textrm{MeV}$, only the real
part contributes. At the energy near $2536 \ \textrm{MeV}$, both the
real and imaginary parts have contributions. We also vary $\Lambda$
to see the cutoff dependence and find that the real part is
sensitive to $\Lambda$, while the imaginary part is unchanged.

\begin{figure}[htbp]
\begin{minipage}[l]{0.45\textwidth}
  \centering
  \includegraphics[scale=0.7]{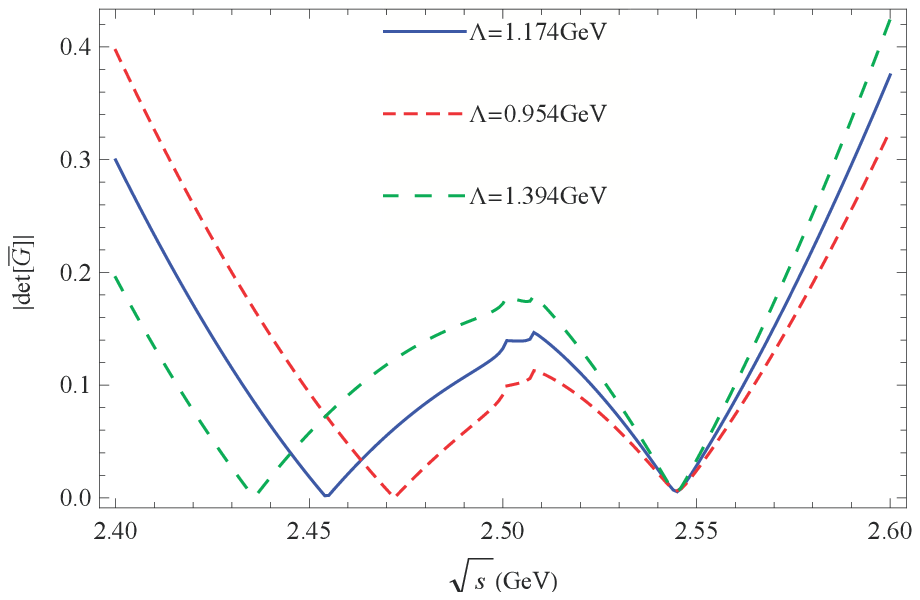}\\
  \vspace{-0.3cm}
  \caption{Pole structures highlighted by the zero values of $\det[\bar{G}]$ in the propagator matrix. }
  \label{fig8}.
\end{minipage} \ \ \
\begin{minipage}[l]{0.45\textwidth}
  \centering
  \vspace{-0.4cm}
  \includegraphics[scale=0.8]{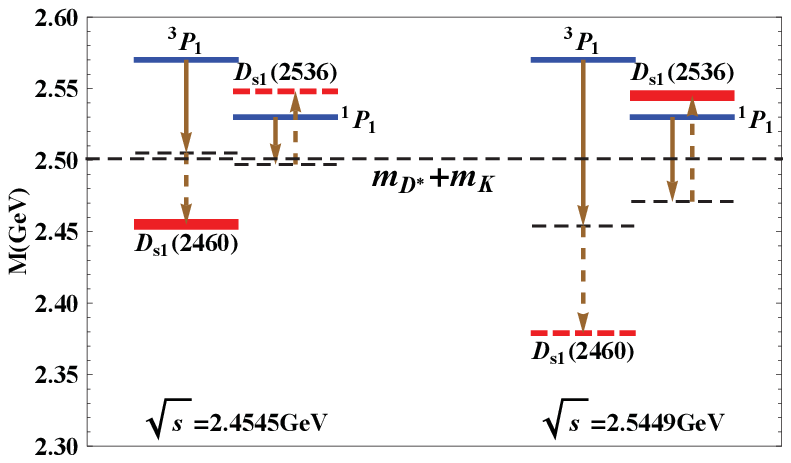}\\
  \vspace{0.22cm}
  \caption{Schematic plot for the mass-shift procedure~\cite{Wu:2011yb}.}
  \label{fig9}
\end{minipage}
\end{figure}

The poles in the propagator matrix $G_{ab}^{\mu \nu}$ are equal to
the zero points in $\bar{G}_{ab}$ as shown in Fig.~\ref{fig8}. For
the bare $c\bar{s}$ masses, we adopt the values from the GI model,
$m[{}^3P_1]=2.57 \ \textrm{GeV}$ and $m[{}^1P_1]=2.53 \
\textrm{GeV}$~\cite{Godfrey:1985xj}. From Fig.~\ref{fig8} we can see
two poles: one is near $2460 \ \textrm{MeV}$ and the other is near
$2536 \ \textrm{MeV}$. The higher pole is insensitive to the cutoff
while the lower pole turns out to be rather sensitive.

We list the masses and widths in Table~\ref{tab:mass} with
$\Lambda=1.174 \ \textrm{GeV}$. The results agree with experiment,
which indicate that the exponential form factor from the quark model
can give a good estimate of the real part. The detailed mass shift
procedures of these two physical states are demonstrated in
Fig.~\ref{fig9}.
\begin{table}[htbp]
  \centering
  \caption{Masses and widths obtained from the pole analysis~\cite{Wu:2011yb}.}\label{tab:mass}
  \begin{tabular}{c|c|c}
    \hline \hline
    $[m-i\frac{\Gamma}{2}]$ (MeV) & $D_{s1}(2460)$ & $D_{s1}(2536)$ \\ \hline
    $\Pi_{ab}^1              $ & $2454.5$       & $2544.9-1.0 i$ \\ \hline
    $\Pi_{ab}^1+\Pi_{ab}^2$ & $2455.8$       & $2544.9-1.1 i$ \\ \hline
    Experiment                 & $2459.5$       & $2535.08-0.46 i$ \\
    \hline \hline
  \end{tabular}
\end{table}

Besides the masses and widths, we can also extract the mixing
parameters and the results are shown in Table~\ref{tab:mixing}. We
can see that the mixing angles are different for these two physical
states at their own mass shells. It is also interesting to see the
deviation of the mixing angles from the heavy quark limit. For
$D_{s1}(2460)$ the deviation is large $\Delta\theta=12.3^\circ$, and
for $D_{s1}(2536)$ the deviation is relatively small
$\Delta\theta=4.4^\circ$. The new experimental data from BaBar put a
strong constraint on the mixing angle of $D_{s1}(2536)$, but with
two solutions, i.e. $\theta_1=32.1^\circ$ or
$\theta_2=38.4^\circ$~\cite{Wu:2011yb}, which are symmetric to the
ideal mixing angle. It shows that our theoretical analysis favors
the bigger one.
\begin{table}[htbp]
  \centering
\caption{The mixing angle $\theta$ and relative phase $\phi$
extracted at the two poles in those three diagonalization
schemes~\cite{Wu:2011yb}.}
  \label{tab:mixing}
  \begin{tabular}{|c||c|c|c||c|c|c|}
    \hline
    & \multicolumn{3}{c||}{$D_{s1}(2460)$} & \multicolumn{3}{c|}{$D_{s1}(2536)$} \\ \cline{2-7}
    \raisebox{2.3ex}[0pt]{$\{\theta,\phi\}[{}^{\circ}]$}& \Rmnum{1} & \Rmnum{2} & \Rmnum{3} & \Rmnum{1} & \Rmnum{2} & \Rmnum{3} \\ \hline
    $\Pi_{ab}^1           $ & $\{47.5, \ 0\}$ & $\{47.5, \ 0\}$ & $\{47.5, \ 0\}$ & $\{39.7, \ -6.4\}$ & $\{39.7, \ 6.4\}$ & $\{39.7, \ 0\}$ \\ \hline
    $\Pi_{ab}^1+\Pi_{ab}^2$ & $\{47.6, \ 0\}$ & $\{47.6, \ 0\}$ & $\{47.6, \ 0\}$ & $\{39.8, \ -6.5\}$ & $\{39.8, \ 6.5\}$ & $\{39.7, \  0\}$ \\
    \hline
  \end{tabular}
\end{table}

\section{Summary}

When taking into account the $D^*K$ loop corrections in the
Godfrey-Isgur model, we can explain the masses and widths of the
physical states $D_{s1}(2460)$ and $D_{s1}(2536)$, and extract their
mixing angles. We also find that when there are strong $S$-wave
coupling channels, the hadron loop corrections can cause both large
mass shifts from the quark model prediction and large mixing angle
shifts from the heavy quark limit. Also, from our calculation we can
see that the exponential form factor from the quark model can give a
good estimate of the real part of the meson loops. Our study may
provide further insights into the role of hadron loops as an
important unquenched correction to the potential quark model.

\section{Acknowledgement}

This work is supported, in part, by National Natural Science
Foundation of China (Grant No.  11035006), Chinese Academy of
Sciences (KJCX2-EW-N01), and Ministry of Science and Technology of
China (2009CB825200).

\end{document}